\documentstyle[aps,preprint]{revtex}
\begin{document}
\draft

\newcommand{\pp}[1]{\phantom{#1}}
\newcommand{\uu}[1]{\underline{#1}}
\newcommand{\be}{\begin{eqnarray}}
\newcommand{\ee}{\end{eqnarray}}
\newcommand{\ve}{\varepsilon}
\newcommand{\vs}{\varsigma}
\newcommand{\Tr}{{\,\rm Tr\,}}
\newcommand{\pol}{\frac{1}{2}}

\title{
A toy model of bosonic non-canonical quantum field
}
\author{Marek~Czachor and Monika Syty}
\address{
Katedra Fizyki Teoretycznej i Metod Matematycznych\\
Politechnika Gda\'{n}ska,
ul. Narutowicza 11/12, 80-952 Gda\'{n}sk, Poland}
\maketitle

\begin{abstract}
A harmonic oscillator is an indefinite-frequency one if the parameter
$\omega$ is replaced by an operator. An ensemble of $N$ such oscillators
may be regarded as a toy model of a bosonic quantum field. All the
possible frequencies associated with a given problem are present
already in a single oscillator and $N$ can be finite. Due to the
operator character of $\omega$ the resulting algebra of
creation-annihilation operators is non-canonical.  In the limit of
large $N$ one recovers perturbation 
theory formulas of the canonical
quantum field theory but with form factors automatically built in.
Vacuum energy of the ensemble is finite, a fact discussed in the
context of the cosmological constant problem. Space of states is
given by a vector bundle with Fock-type fibers. Interactions of the
field with 2-level systems, including Rabi oscillations and
spontaneous emission, are discussed in detail. 
\end{abstract}

%\narrowtext

\section{Introduction}

It is well known that standard canonical procedures of field
quantization result in various infinities which have to be removed
in a more or less ad hoc manner. In
spite of unquestionable successes of quantum field theories, there
exists a possibility that we are still overlooking an ingredient
which is essential for a physically consistent field quantization. 

Quite recently it was pointed out \cite{I} that the very first step
of field quantization may be performed incorrectly. The point is that
in the description of quantum harmonic oscillators {\it \`a la\/}
Heisenberg one treats the frequency $\omega$ in 
\be
H_\omega=\frac{\hat p^2}{2m}+\frac{m\omega^2 \hat q^2}{2}\label{1}
\ee
as a parameter. A mathematical implication of this fact is the
algebra of canonical commutation relations (CCR) characterizing
creation and annihiltion operators. 

However, thinking of real systems (such as a simple 
pendulum, or a string) one realizes that $\omega$ typically depends on other 
physical quantities (say, positions) which in quantum theories 
are represented by operators. As a 
consequence it is not entirely irrelevant to ask what would be changed in the 
theory if one quantized not only $p$ and $q$ but also $\omega$ 
\cite{footnote}. 

The main physical conclusion of the analysis given in \cite{I} is that at
least vacuum and ultraviolet infinities may disappear as a result of
this single modification \cite{infrared}. Mathematically the effect is
rooted in non-canonical comutation relations
(non-CCR) which naturally replace CCR. 

The purpose of the present paper is to take a closer look at the
meaning of the main assumption of \cite{I}. We return to the first
step of the construction and concentrate on a nonrelativistic
harmonic oscillator. We replace $\omega$ by an operator. We
discuss in detail creation and annihilation operators and
modifications of CCR. We arrive at
the same non-CCR algebra as in \cite{I} but now in a
slightly modified representation. Evaluating averages of position and
momentum we arrive at 
expressions resembling Fourier expansions of classical fields. 

The next step is to discuss interactions of such indefinite-frequency
oscillators with two-level systems. We study the excited-state
survival amplitude for a two-level system interacting with $N$
oscillators and obtain an exact formula in terms of Laplace
transforms.  

For any $N$ and $\omega$ with
only one frequency we obtain the Rabi oscillation. A more realistic
case is when the set of all the $\omega$s corresponds to a cavity
spectrum. One expects Rabi oscilations also if one frequency is in
resonance with the two-level system while the remaining $\omega$s are
very far from resonance. For small $N$ an appropriate probability
differs essentially from the Jaynes-Cummings solution. However, we
show that for large $N$ one expects a solution analogous to the
standard one.  

We next show that to any
order of perturbation theory and to any given precision
the perturbative predictions of the canonical theory may be reconstructed 
in the non-canonical formalism with a {\it finite\/} number of
oscillators, and the ``canonical infinities" do not
occur. We thus generalize the results from \cite{I} which were shown
explicitly only in low orders of perturbation theory.

Since in the non-canonical formalism the main feature is an automatic
elimination of divergent expressions, it is interesting to discuss
implications of finite vacuum contribution to the cosmological
constant problem. We show that in the simplest approach the well
known formula for the vacuum energy density, 
$
\rho_{\rm old}
\approx
\frac{\hbar}{16\pi^2c^3}\omega_{\rm max}^4
$
is replaced by 
$\rho_{\rm new}
\approx
\frac{3}{8}\hbar\omega_{\rm max}\frac{\langle N\rangle}{V}
$
where $\langle N\rangle/V$ is the average number of oscillators in
volume $V$. The physical meaning of the cut-off frequency is also
different from the usual one: This is not the cut-off in the energy
spectrum, but the frequency above which the vacuum probability
density of $\omega$'s is negligible. The very existence of such a
parameter in non-canonical theories trivially follows from square
integrability of vacuum wave functions. Experimental values of vacuum
energy density allow to estimate the average number of oscillators in
a unit volume. The resulting number seems too small to guarantee
consistency of non-canonical quantum optics with atomic measurements,
which suggests that before any comparison of non-canonical vacuum
contributions with cosmology one should first discuss non-canonical
theories with broken supersymmetry.  

It should be stressed that what we are doing in the paper is not,
strictly speaking, a new field quantization. We simply apply orthodox
quantum mechanics to systems of many indefinite-frequency harmonic
oscillators. The point we advocate is that for {\it any\/} number 
$N$ of such oscillators (also $N<\infty$) the resulting ensemble may
be regarded as a model for a spin-zero quantum field. 
In the limit $N\to \infty$ the ensemble has
properties of the standard canonical quantum field, but in a version
with form-factors automatically built in. Quantum field theoretic
interpretation of the ensemble requires an appropriate construction
of the space of states. We show that the non-CCR state space is not a
Fock one, but rather a vector bundle with the set of vacua in the
role of a base space and Fock spaces playing the role of fibers.  

\section{Indefinite-frequency oscillator}

We begin with (\ref{1})
where $[\hat q,\hat p]=i\hbar \hat 1$ and $\omega$ is a parameter. 
Wavefunctions
related to  $H_\omega$
are, in position representation, functions of $q$ i.e. $\psi=\psi(q)$
with the normalization 
$\int_{-\infty}^\infty dq|\psi(q)|^2=1$. 
Now regard $\omega$ as a {\it quantum number\/}, i.e. an eigenvalue
of some operator $\Omega$. This means that the
wavefunctions depend on two parameters, $\psi=\psi(\omega,q)$, and are
normalized by 
$$\int_{-\infty}^\infty dq\int_0^\infty d\omega|\psi(\omega,q)|^2=1,$$
if spectrum of $\Omega$ is continuous, or by 
$$\sum_\omega \int_{-\infty}^\infty dq|\psi(\omega,q)|^2=1,$$
if spectrum is discrete \cite{footnote}.

\subsection{Algebra of indefinite-frequency operators}

At the level of observables we can formalize this by
means of the operators
\be
P &=& \hat 1\otimes \hat p\\
Q &=& \hat 1\otimes \hat q\\
\Omega &=& \hat \omega \otimes \hat 1\\
H &=&\frac{P^2}{2m}+\frac{m\Omega^2 Q^2}{2}
\ee
whose explicit representation is
\be
P \psi(\omega,q) &=& -i\hbar
\frac{\partial\psi(\omega,q)}{\partial q}\\
Q \psi(\omega,q) &=& q\psi(\omega,q)\\
\Omega \psi(\omega,q) &=& \omega\psi(\omega,q)\\
H \psi(\omega,q)
&=&
\Big(-\frac{\hbar^2}{2m}\frac{\partial^2}{\partial q^2}
+
\frac{m\omega^2 q^2}{2}\Big)\psi(\omega,q).\label{hat H}
\ee
$[Q,P] =  i\hbar I$ where $I=\hat 1\otimes \hat 1$ \cite{hat 1}. 
As we can see no drastic changes were made so far by the reinterpretation of
$\omega$.  

The creation and annihilation operators are defined via an obvious
generalization of the standard formulas
\be
a{_\Omega}
&=&
\sqrt{\frac{m\Omega}{2\hbar}}Q+i
\sqrt{\frac{1}{2m\hbar\Omega}}P
\\
a{_\Omega}^{\dag}
&=&
\sqrt{\frac{m\Omega}{2\hbar}}Q-i
\sqrt{\frac{1}{2m\hbar\Omega}}P\\
I &=& {[a{_\Omega},a{_\Omega}^{\dag}]}
\ee
and satisfy 
\be
H
&=&
\frac{\hbar\Omega}{2}\Big(a{_\Omega}^{\dag}a{_\Omega}+
a{_\Omega} a{_\Omega}^{\dag}\Big)
\ee
Assume for simplicity that spectrum of $\Omega$ is discrete, i.e. its
spectral representation reads
\be
\Omega
&=&
\sum_\omega \omega |\omega\rangle\langle \omega|\otimes \hat 1.
\ee
The Hamiltonian can be now written as 
\be
H
&=&
\sum_\omega\frac{\hbar\omega}{2}\Big(a_\omega^{\dag}a_\omega+
a_\omega a_\omega^{\dag}\Big)
\ee
where 
\be
a_\omega
&=&
\sqrt{\frac{m\omega}{2\hbar}}|\omega\rangle\langle
\omega|\otimes \hat q+i
\sqrt{\frac{1}{2m\hbar\omega}}|\omega\rangle\langle \omega|\otimes
\hat p\\ 
&=&
|\omega\rangle\langle
\omega|\otimes \Big(
\sqrt{\frac{m\omega}{2\hbar}}\hat q+i
\sqrt{\frac{1}{2m\hbar\omega}}
\hat p\Big)\\
&=&
|\omega\rangle\langle
\omega|\otimes \hat a_\omega.
\ee
Here $\hat a_\omega$ is the standard annihilation operator for an
oscillator with frequency given by the {\it parameter\/} $\omega$. 
$\hat a_\omega$ satisfies canonical commutation relations (CCR)
\be
[\hat a_\omega,\hat a_\omega^{\dag}]=\hat 1.
\ee
In the standard formalism one does not ask for the commutator 
$[\hat a_\omega,\hat a_{\omega'}^{\dag}]$ since a single oscillator
has only one frequency parameter and the question is physically ill posed. 
Two different $\omega$ and $\omega'$ can
occur only if one has two different oscillators and then the
commutator vanishes. 

The algebra of $a_\omega$ is non-canonical (non-CCR):
\be
[a_\omega,a_{\omega'}^{\dag}]=\delta_{\omega\omega'}|\omega\rangle\langle
\omega|\otimes \hat 1=:\delta_{\omega\omega'}I_\omega \label{nccr}.
\ee
Now the commutator corresponds to a well posed physical problem.

A notable property of the non-CCR algebra (\ref{nccr}) is the resolution
of identity satisfied by the right-hand-side of the commutator:
\be
\sum_\omega I_\omega=I.\label{resolution}
\ee

\subsection{Indefinite-frequency states}

Eq. (\ref{hat H}) shows that energy eigenvector corresponding to the
eigenvalue
\be
E(\omega,n)=\hbar\omega (n+\textstyle{\frac{1}{2}})
\ee
is 
\be
|\omega,n\rangle
=
|\omega\rangle|n_\omega\rangle.\label{basis}
\ee
By $|n_\omega\rangle$ we denote the basis associated with the
decomposition 
\be
\hat a_\omega=\sum_{n=0}^\infty \sqrt{n+1}|n_\omega\rangle 
\langle n+1_\omega|
\ee
of the CCR annihilation operator. The associated position-space
wavefunction is 
\be
\langle q|n_\omega\rangle
\sim e^{-q^2m\omega/(2\hbar)}h_n(q\sqrt{m \omega/(2\hbar)})
\ee
with $h_n(\cdot)$ the Hermite polynomial. 

One of the properties that make non-CCR oscillator interesting is
that the average energy evaluated in the state 
\be
|\psi\rangle=\sum_{\omega,n}\psi(\omega,n)|\omega,n\rangle
\ee
can be written as 
\be
\langle \psi|H|\psi \rangle
&=& 
\sum_{\omega,n} \hbar\omega(n+\textstyle{\frac{1}{2}})|\psi(\omega,n)|^2
\ee
i.e. in a form which looks like an average energy of an ensemble of
independent oscillators with different frequencies: A {\it single\/}
indefinite-frequency oscillator in many respects resembles an {\it
ensemble\/} of many independent oscillators.  

We define {\it a\/} vacuum state as {\it any\/} state $|O\rangle$ which is
annihilated by $a{_\Omega}$ i.e. 
\be
|O\rangle
&=&
\sum_\omega O(\omega)|\omega,0\rangle.
\ee
Let us note that a general vacuum state can be time dependent. 
In particular 
\be
|O_t\rangle
&=&
e^{i \Omega t/2}|O\rangle \label{v pic}
\ee
is also a vacuum state. The unitary transformation (\ref{v pic})
defines a ``vacuum picture" whose dynamics is given by the
Hamiltonian 
\be
\tilde H
&=&
\hbar\Omega a{_\Omega}^{\dag}a{_\Omega}
\ee
Having {\it a\/} vacuum we can define $N$th excited
states by 
\be
|N\rangle
&=&
\frac{1}{\sqrt{N!}}(a{_\Omega}^{\dag})^N|O\rangle.
\ee
$|N\rangle$ represents a superposition of oscillators with
different frequencies but the same level of excitation. 

Coherent states can be defined in the usual way via the displacement operator
\be
D(z)&=&\exp\big( z a{_\Omega}^{\dag}-\bar z a{_\Omega}\big)\\
|z\rangle
&=&
D(z)|O\rangle\\
a{_\Omega}|z\rangle
&=&
z|z\rangle
\ee
One can further generalize coherent states by taking any operator function 
\be
f{_\Omega}=
\sum_\omega f(\omega) |\omega\rangle\langle \omega|\otimes \hat 1.
\ee
and the displacement operator 
\be
D(f{_\Omega})&=&\exp\big( f{_\Omega} a{_\Omega}^{\dag}-f{_\Omega}^{\dag} 
a{_\Omega}\big)\\
|f{_\Omega}\rangle
&=&
D(f{_\Omega})|O\rangle\\
&=&\sum_\omega O(\omega) 
e^{f(\omega) a_\omega^{\dag}-\overline{f(\omega)}
a_\omega}|\omega,0\rangle\\
&=&\sum_\omega O(\omega) 
|\omega\rangle e^{f(\omega) \hat a_\omega^{\dag}-\overline{f(\omega)}
\hat a_\omega}|0_\omega\rangle\\
&=&\sum_\omega O(\omega) 
|\omega\rangle|f(\omega)_\omega\rangle\\
a{_\Omega}|f{_\Omega}\rangle
&=&
f{_\Omega}|f{_\Omega}\rangle
=
\sum_\omega O(\omega)f(\omega) 
|\omega\rangle|f(\omega)_\omega\rangle\label{eigen}
\ee
Eq. (\ref{eigen}) means that coherent states are generalized 
eigenstates of annihilation operators.

\subsection{Single-oscillator ``fields"}

The Heisenberg picture dynamics is given by the familiar formulas 
\be
e^{iHt/\hbar}a{_\Omega} e^{-iHt/\hbar}
&=&
e^{-i\Omega t}a{_\Omega}
=
e^{i\tilde Ht/\hbar}a{_\Omega} e^{-i\tilde Ht/\hbar}
,\\
e^{iHt/\hbar}a_\omega e^{-iHt/\hbar}
&=&
e^{-i \omega t}a_\omega
=
e^{i\tilde Ht/\hbar}a_\omega e^{-i\tilde Ht/\hbar}
\ee
implying 
\be
Q_t=e^{iHt/\hbar}Q e^{-iHt/\hbar}
&=&
\sqrt{\frac{\hbar}{2\Omega m}}\Big(
e^{-i\Omega t}a{_\Omega}
+
e^{i\Omega t}a{_\Omega}^{\dag}\Big)
\\
P_t=e^{iHt/\hbar}P e^{-iHt/\hbar}
&=&
-i\sqrt{\frac{\hbar \Omega m}{2}}\Big(
e^{-i\Omega t}a{_\Omega}
-
e^{i\Omega t}a{_\Omega}^{\dag}\Big)
\ee
Evaluating the coherent-state averages 
\be
\langle f{_\Omega}| Q_t|f{_\Omega}\rangle
&=&
\sum_\omega |O(\omega)|^2 \sqrt{\frac{\hbar}{2\omega m}}\Big(
e^{-i\omega t}f(\omega)
+
e^{i\omega t}\overline{f(\omega)}\Big)
\\
\langle f{_\Omega}| P_t|f{_\Omega}\rangle
&=&
-i\sum_\omega |O(\omega)|^2
\sqrt{\frac{\hbar \omega m}{2}}\Big(
e^{-i\omega t}f(\omega)
-
e^{i\omega t}\overline{f(\omega)}\Big)
\ee
we realize that a single harmonic oscillator with indefinite
frequency is an object closely related to quantum fields. 

\section{Many oscillators}

An ensemble of many indefinite-frequency oscillators has properties
analogous to a quantum field. 
Formally the multi-oscillator structure is constructed as follows.

Let $A$ be an operator at a ``single-oscillator" level. If ${\cal H}$
is the Hilbert space of one-oscillator states then 
$A: {\cal H} \to {\cal H}$. Let 
\be
\uu {\cal H}=\oplus_{n=1}^\infty \otimes_s^n {\cal H}
\ee
be the Hilbert space of states corresponding to an indefinite number
of oscillators; $\otimes_s^n {\cal H}$ stands for a space of
symmetric states in $\underbrace{{\cal H}\otimes\dots \otimes {\cal H}}_n$.
We introduce the following notation for operators defined at the
multi-oscillator level:
\be
\oplus_{\alpha_n} A
&=&
\alpha_1 A
\oplus 
\alpha_2\big(A\otimes I+I\otimes A\big)\oplus
\alpha_3\big(A\otimes I\otimes I+
I\otimes A \otimes I+
I\otimes I\otimes A\big)\oplus
\dots
\ee
Here $\oplus_{\alpha_n} A: \uu {\cal H}\to \uu {\cal H}$ and
$\alpha_n$ are real or complex parameters. 

The following properties follow directly from the definition 
\be
[\oplus_{\alpha_n} A,\oplus_{\beta_n} B]&=&\oplus_{\alpha_n\beta_n}
[A,B]\\
e^{\oplus_{\alpha_n} A}
&=&
\oplus_{n=1}^\infty
\underbrace{e^{\alpha_n A}\otimes\dots \otimes e^{\alpha_n A}}_n\\
e^{\oplus_1 A}\oplus_{\beta_n} Be^{-\oplus_1 A}
&=&
\oplus_{\beta_n}e^{A}Be^{-A}
\ee
%bbbbbbbbbbbbbbbbbb
Identity operators  at $\cal H$ and $\uu {\cal H}$ are related by 
\be
\uu I &=& \oplus_{\frac{1}{n}} I.
\ee 
Define multi-oscillator creation and annihilation operators by 
\be
\uu a{_\Omega} &=&\oplus_{\frac{1}{\sqrt{n}}} a{_\Omega},\label{apocz}\\
\uu a{_\Omega}^{\dag} &=&\oplus_{\frac{1}{\sqrt{n}}}
a{_\Omega}^{\dag}\\
\uu a{_\omega} &=&\oplus_{\frac{1}{\sqrt{n}}} a{_\omega},\\
\uu a{_\omega}^{\dag} &=&\oplus_{\frac{1}{\sqrt{n}}}
a{_\omega}^{\dag}.\label{akonc}
\ee
Then 
\be
[\uu a{_\Omega},\uu a{_\Omega}^{\dag}]
&=&
\oplus_{\frac{1}{n}} [a{_\Omega},a{_\Omega}^{\dag}]=
\oplus_{\frac{1}{n}} I=\uu I\label{uu ccr}\\
{[\uu a{_\omega},\uu a{_{\omega'}}^{\dag}]}
&=&
\oplus_{\frac{1}{n}} [a{_\omega},a{_{\omega'}}^{\dag}]=
\oplus_{\frac{1}{n}}\delta_{\omega\omega'} I_\omega=:
\delta_{\omega\omega'}\uu I_\omega\\
\ee
The apparently artificial factor $1/\sqrt{n}$ in
(\ref{apocz})--(\ref{akonc}) is needed to maintain the CCR condition
(\ref{uu ccr}). Another consequence of this choice of $\alpha_n$ is
the resolution of identity 
\be
\sum_\omega \uu I_\omega=\uu I
\ee
which is the multi-oscillator counterpart of (\ref{resolution}). 
The Hamiltonian of an ensemble of noninteracting oscillators is 
\be
\uu H
&=&
\oplus_1 H\\
&=&
\oplus_1 
\frac{\hbar\Omega}{2}
\Big(a{_\Omega}a{_\Omega}^{\dag}+a{_\Omega}^{\dag}a{_\Omega}\Big)
\\
&=&
\oplus_1 \Big(\frac{P^2}{2m}+\frac{m\Omega^2 Q^2}{2}\Big)
\ee
This is the standard form of Hamiltonian corresponding to many
noninteracting particles.  

Denoting $\uu \Omega=\oplus_1\Omega$ and proceeding 
similarly to the single-oscillator case we can define the vacuum
picture by means of the unitary transformation 
\be
|\tilde {\uu\psi}\rangle=e^{i\uu \Omega t/2}|\uu \psi\rangle
\ee
and the vacuum-picture Hamiltonian is 
\be
\uu {\tilde H}
&=&
\oplus_1 \tilde H
=
\oplus_1 
\hbar\Omega
a{_\Omega}^{\dag}a{_\Omega}.
\ee
There is a subtle difference between the way we introduce the vacuum
picture and the standard way of removing the infinite energy of
vacuum. In our case the operation is given by a well defined unitary
operator, whereas the standard procedure involves a ``phase factor"
$e^{i\infty t}$ which, in a strict mathematical sense, does not
exist.  

\section{Jaynes-Cummings interaction}

Although the above oscillator is formally a ``single-mode" one we
have seen that the coherent-state averages of $Q_t$ and $P_t$
resemble Fourier 
decompositions of classical fields. 

A single-mode interaction of an oscillator with a two-level
system can be solved analytically in the rotating wave approximation
(RWA). One knows that, in the standard formalism, oscillations with
Rabi frequencies will occur. In our case one expects a {\it
superposition\/} of different Rabi frequencies and hence a
possibility of irreversible spontaneous emission is not excluded. 

For a single harmonic oscillator one can repeat standard Heisenberg
picture calculations described in detail in \cite{Allen}. We found it
more instructive to begin with the more general case of an arbitrary
(or indefinite) number of oscillators. Even at such a general level
the problem can be exactly solved. 

For a single oscillator the vacuum-picture Hamiltonian is
\be
H_{\rm tot}
&=&
\hbar\omega_0 R_3 +\tilde H+\hbar\alpha R_2 Q
\\
&=&
\hbar\omega_0 R_3 +\tilde H+\hbar\alpha R_2
\sqrt{\frac{\hbar}{2\Omega m}}\Big(
a{_\Omega}
+
a{_\Omega}^{\dag}\Big)
\ee
and its RWA version reads
\be
H_{\rm rwa}&=&
\hbar\omega_0 R_3 +\hbar\Omega a{_\Omega}^{\dag}a{_\Omega}
+i\frac{\hbar\alpha}{2} 
\sqrt{\frac{\hbar}{2\Omega m}}\Big(
R_+a{_\Omega}
-
R_-a{_\Omega}^{\dag}\Big)
\label{RWA}
\ee
where $R_k=\sigma_k/2$, $R_\pm=R_1\pm iR_2$, and $\alpha$ is a real constant. 
The frequency operator $\Omega$ commutes with $R_k$,
$a{_\Omega}$, $a{_\Omega}^{\dag}$, and $H$. 

The interaction term of the model describing interaction of a
two-level system with one oscillator is $\hbar\alpha R_2 Q$. Here $Q$
is the operator representing a configuration-space position of the
oscillator. Having a single two-level system interacting with, say, two
independent oscillators one expects a term of the form 
\be
\hbar\alpha_2 R_2 Q_1+
\hbar\alpha_2 R_2 Q_2
\ee
where $Q_k$ commute with each other (at equal times). 
The coupling constant may depend on the number
of oscillators. A classical intuition suggests that the greater
number of oscillators crowding around the two-level system, the
weaker the interaction of a single element of the ensemble. 
A similar property 
of coupling constants is found in rigorous approaches to
thermodynamic limit in Bose-Einstein condensates \cite{Lieb}. 

Several different ways of reasoning \cite{I} suggest
$\alpha_n=\alpha/\sqrt{n}$ which implies (in the vacuum picture) 
\be
\uu H_{\rm tot}
&=&
\hbar\omega_0 R_3 +\uu{\tilde H}+\hbar R_2 
\big[\oplus_\frac{\alpha}{\sqrt{n}} Q\big]
\\
&=&
\hbar\omega_0 R_3 +\oplus_1\hbar \Omega a{_\Omega}^{\dag}a{_\Omega}
+\hbar\alpha R_2
\Big[\oplus_\frac{1}{\sqrt{n}}\sqrt{\frac{\hbar}{2\Omega m}}\Big(
a{_\Omega}
+
a{_\Omega}^{\dag}\Big)\Big]\\
&=&
\hbar\omega_0 R_3 +\sum_\omega\oplus_1\hbar \omega a{_\omega}^{\dag}a{_\omega}
+\hbar\alpha R_2\sum_\omega
\sqrt{\frac{\hbar}{2\omega m}}\Big(
\uu a{_\omega}
+
\uu a{_\omega}^{\dag}\Big)
\ee
The RWA
Hamiltonian is now 
\be
\uu H_{\rm rwa}
&=&
\hbar\omega_0 R_3 +\uu {\tilde H}+i\frac{\hbar\alpha}{2}\sum_\omega
\sqrt{\frac{\hbar}{2\omega m}}\Big(
R_+\uu a{_\omega}
-
R_-\uu a{_\omega}^{\dag}\Big)\\
&=&
\hbar\omega_0 R_3 +\oplus_1 \hbar\Omega
a{_\Omega}^{\dag}a{_\Omega}
+
i\frac{\hbar\alpha}{2}
\Big(
R_+\Big[\oplus_\frac{1}{\sqrt{n}}\sqrt{\frac{\hbar}{2\Omega m}} 
a{_\Omega}\Big]
-
R_-\Big[\oplus_\frac{1}{\sqrt{n}}\sqrt{\frac{\hbar}{2\Omega m}} 
a{_\Omega}^{\dag}\Big]\Big)
\ee
It is clear that our indefinite-frequency oscillators may be regarded
as a model of a scalar quantum field interacting with a 2-level atom
located at the origin. 

In order to compute the evolution of atomic inversion we will proceed
in two ways.  
We will begin with a Dyson expansion and then derive an integral
equation which will be solved by means of Laplace transforms. Both
methods are instructive and show in different ways links to the
standard canonical theory.  

We start with
\be
\uu H_0
&=&
\hbar\omega_0 R_3 +\oplus_1 \hbar\Omega
a{_\Omega}^{\dag}a{_\Omega}\\
\uu H_1
&=&
i\frac{\hbar\alpha}{2}
\Big(
R_+\Big[\oplus_\frac{1}{\sqrt{n}}\sqrt{\frac{\hbar}{2\Omega m}} 
a{_\Omega}\Big]
-
R_-\Big[\oplus_\frac{1}{\sqrt{n}}\sqrt{\frac{\hbar}{2\Omega m}} 
a{_\Omega}^{\dag}\Big]\Big)
\ee
and 
\be
\uu H_{1\rm int}(t)
&=&
e^{i\uu H_0t/\hbar}\uu H_1e^{-i\uu H_0t/\hbar}\\
&=&
i\frac{\hbar\alpha}{2}
\Big(
e^{i\omega_0 t}R_+
\Big[\oplus_\frac{1}{\sqrt{n}}\sqrt{\frac{\hbar}{2\Omega m}} 
e^{-i\Omega t}a{_\Omega}\Big]
-
e^{-i\omega_0 t}R_-\Big[\oplus_\frac{1}{\sqrt{n}}
\sqrt{\frac{\hbar}{2\Omega m}} 
e^{i\Omega t}a{_\Omega}^{\dag}\Big]\Big)\\
&=&
i\frac{\hbar\alpha}{2}
\sum_\omega\Big(
R_+
\sqrt{\frac{\hbar}{2\omega m}} 
e^{-i(\omega-\omega_0) t}\uu a{_\omega}
-
R_-
\sqrt{\frac{\hbar}{2\omega m}} 
e^{i(\omega-\omega_0) t}\uu a{_\omega}^{\dag}\Big)
\ee
\section{Survival probability}

Survival probability is the probability that at $t>0$ no spontaneous
emission occured. In the context of our model this means one begins
at $t=0$ with the state 
\be
|\uu O\rangle
&=&
\sum_{\omega}O^{(1)}(\omega)|\omega,0\rangle|+\rangle\nonumber\\
&\pp =&\oplus
\sum_{\omega_1,\omega_2}
O^{(2)}(\omega_1,\omega_2)|\omega_1,0\rangle
|\omega_2,0\rangle|+\rangle\nonumber\\
&\pp =&\oplus\dots
\ee
representing all the oscillators in their ground states and the
2-level system in the excited state. The projector on the subspace
consisting of all such states is 
\be
P
&=&
\oplus_{N=1}^\infty \sum_{\omega_1\dots\omega_N}
|\omega_1,0\rangle\langle\omega_1,0|\otimes \dots\otimes 
|\omega_N,0\rangle\langle\omega_N,0|\otimes
|+\rangle\langle+|
\ee
This should be contrasted with the standard CCR case where there is
only one vacuum $|0_{\rm ccr}\rangle$, the initial state is 
\be
|O_{\rm ccr}\rangle
=
|0_{\rm ccr}\rangle|+\rangle
\ee
and the projector is on a one-dimensional subspace,
\be
P_{\rm ccr}
=
|O_{\rm ccr}\rangle
\langle O_{\rm ccr}|
\ee
Denoting, as before, by $\uu H$ and $\uu {\tilde H}$ the Hamiltonians
in Schr\"odinger and vacuum pictures, respectively, we can consider
solutions of 
\be
i\hbar |\dot {\uu\psi}\rangle
=
\uu H |{\uu\psi}\rangle
\ee
and 
\be
i\hbar |\dot {\uu{\tilde\psi}}\rangle
=
\uu {\tilde H} |{\uu{\tilde \psi}}\rangle
\ee
which have the same initial condition $|\uu O\rangle$
at $t=0$. There are now several different but physically natural
``survival probabilities": 
\be
|\langle \uu O| e^{-i\uu Ht/\hbar}|\uu O\rangle|^2 \label{sp1},
\ee
\be
|\langle \uu O| e^{-i\uu {\tilde H}t/\hbar}|\uu O\rangle|^2 \label{sp2}
\ee
and 
\be
\langle \uu O| e^{i\uu Ht/\hbar}Pe^{-i\uu Ht/\hbar}|\uu O\rangle 
=
\langle \uu O| e^{i\uu {\tilde H}t/\hbar}Pe^{-i\uu {\tilde
H}t/\hbar}|\uu O\rangle 
\label{sp3},
\ee
(\ref{sp1}) cannot correspond to the probability we want to calculate
since even in the absence of interactions (i.e. when $\alpha=0$) one
finds 
\be
|\langle \uu O| e^{-i\uu Ht/\hbar}|\uu O\rangle|^2
=
|\langle \uu O| e^{-i\uu \Omega t/2}|\uu O\rangle|^2\neq 1
\ee
for $t>0$. Probabilities (\ref{sp2}) and (\ref{sp3}) equal 1 for any
$t$ if $\alpha=0$ but are unequal for $\alpha\neq 0$.  

In the canonical theory the first of these probabilities is not well
defined due to the infinite vacuum energy, but the remaining two are
equal since 
\be
\langle \uu O_{\rm ccr} | e^{i\uu {\tilde H}t/\hbar}P_{\rm
ccr}e^{-i\uu {\tilde H}t/\hbar}|\uu O_{\rm ccr} \rangle 
=
|\langle \uu O_{\rm ccr} |e^{-i\uu {\tilde H}t/\hbar}|\uu O_{\rm ccr}
\rangle|^2, 
\ee
a consequence of one-dimensionality of the vacuum subspace.

In our non-canonical theory the probability which tends (after
renormalization of $\alpha$ and for $N\to\infty$) to the canonical
result is (\ref{sp2}), as we shall see below.  

The vectors 
\be
|\bbox 0_{\omega_1\dots\omega_N}\rangle
&=&
|\omega_1,0\rangle\dots |\omega_N,0\rangle|+\rangle
\\
|\bbox 1_{\omega_1\dots\omega_N}\rangle
&=&
|\omega_1,1\rangle\dots |\omega_N,0\rangle|-\rangle
\\
&\vdots& \nonumber\\
|\bbox N_{\omega_1\dots\omega_N}\rangle
&=&
|\omega_1,0\rangle\dots |\omega_N,1\rangle|-\rangle
\ee
span an invariant subspace with respect to the dynamics.
The subspace contains at most one oscillator in the first excited
state.

Interaction-picture Hamiltonian acts on these vectors as follows 
\be
\uu H_{1\rm int}(t)
|\bbox 0_{\omega_1\dots\omega_N}\rangle
&=&
-
i\frac{1}{\sqrt{N}}\frac{\hbar\alpha}{2}
\sum_{k=1}^N
\sqrt{\frac{\hbar}{2\omega_k m}} 
e^{-i(\omega_0-\omega_k) t}
|\bbox k_{\omega_1\dots\omega_N}\rangle\\
\uu H_{1\rm int}(t)
|\bbox 1_{\omega_1\dots\omega_N}\rangle
&=&
i\frac{1}{\sqrt{N}}\frac{\hbar\alpha}{2}
\sqrt{\frac{\hbar}{2\omega_1 m}} 
e^{i(\omega_0-\omega_1) t}
|\bbox 0_{\omega_1\dots\omega_N}\rangle\\
&\vdots&\nonumber\\
\uu H_{1\rm int}(t)
|\bbox N_{\omega_1\dots\omega_N}\rangle
&=&
i\frac{1}{\sqrt{N}}\frac{\hbar\alpha}{2}
\sqrt{\frac{\hbar}{2\omega_N m}} 
e^{i(\omega_0-\omega_N) t}
|\bbox 0_{\omega_1\dots\omega_N}\rangle
\ee
and can be represented by the matrix
\be
H_N(t)
&=&
\frac{\hbar}{2}
\left(
\begin{array}{cccc}
0 &
-i\frac{\alpha}{\sqrt{N}}
\sqrt{\frac{\hbar}{2\omega_1 m}} e^{-i \Delta_{\omega_1} t}
&
\dots
&
-i\frac{\alpha}{\sqrt{N}}
\sqrt{\frac{\hbar}{2\omega_N m}}e^{-i \Delta_{\omega_N} t} \\
i\frac{\alpha}{\sqrt{N}}
\sqrt{\frac{\hbar}{2\omega_1 m}} e^{i\Delta_{\omega_1} t}
&0&\dots&0\\
\vdots& & & \\
i\frac{\alpha}{\sqrt{N}}
\sqrt{\frac{\hbar}{2\omega_N m}} e^{i\Delta_{\omega_N} t}
&0& \dots &  0
\end{array}
\right)
\ee
where $\Delta_{\omega}=\omega_0-\omega$ are the detunings.
Eigenvectors of this matrix are analogs of dressed states from
the standard formalism. 
To proceed further define two orthonormal vectors:
\be
|1\rangle
&=&
\frac{1}{\sqrt{\frac{1}{\omega_1}+\dots+\frac{1}{\omega_N}}}
\left(
\begin{array}{c}
0 \\
i\sqrt{\frac{1}{\omega_1}}
\\
\vdots\\
i
\sqrt{\frac{1}{\omega_N}}
\end{array}
\right)\\
|0\rangle
&=&
\left(
\begin{array}{c}
1 \\
0
\\
\vdots\\
0
\end{array}
\right)
=|\bbox 0_{\omega_1\dots\omega_N}\rangle\label{zero}
\ee
and the unitary matrix
\be
U_t
&=&
\left(
\begin{array}{cccc}
1 & 0 & \dots & 0\\
0 & e^{i\Delta_{\omega_1} t}&\dots&0\\
\vdots& & & \\
0&0& \dots &  e^{i\Delta_{\omega_N} t}
\end{array}
\right)
\ee
In this notation
\be
H_N(t)
&=&
\frac{\alpha\hbar}{2}
\sqrt{
\frac{\hbar}{2mN}\Big(
\frac{1}{\omega_1}+\dots+\frac{1}{\omega_N}\Big)}
\Big(|0\rangle \langle 1|U_t^{\dag}+U_t|1\rangle \langle 0|
\Big)
\ee
In order to compute (\ref{sp2}) and (\ref{sp3}) it is sufficient to find 
\be
F(t)_{\omega_1\dots\omega_N}
=
\langle 0|e^{-i\uu {\tilde H}t/\hbar}|0\rangle
\ee
Using $\langle 0|U_t|1\rangle=0$, $\langle 0|U_t|0\rangle=1$, and denoting 
\be
f(\tau)=\sum_{k=1}^N\frac{e^{-i\Delta_{\omega_k}\tau}
}{\omega_k}\label{ftau}
\ee
one finds 
\be
F(t)_{\omega_1\dots\omega_N}
&=&
\sum_{n=0}^\infty(-1)^n
\frac{\alpha^{2n}\hbar^n}{(8mN)^n}
\int_0^{t}dt_1\int_0^{t_1}dt_2
\dots
\int_0^{t_{2n-1}}dt_{2n}
f(t_1-t_2)  
\dots
f(t_{2n-1}-t_{2n}) \label{f...f}\\
&=&
1
-
\frac{\alpha^{2}\hbar}{8mN}
\int_0^{t}dt_1\int_0^{t_1}dt_2
f(t_1-t_2) 
F(t_2)_{\omega_1\dots\omega_N}\label{int-eq}
\ee
Differentiating (\ref{int-eq}) with respect to $t$ we get 
\be
\dot F(t)_{\omega_1\dots\omega_N}
=
-\frac{C}{N}
\int_0^{t}dt_2
f(t-t_2) 
F(t_2)_{\omega_1\dots\omega_N},\label{volt}
\ee
$C=\frac{\alpha^{2}\hbar}{8m}$, whose
solution is
\be
F(t)_{\omega_1\dots\omega_N}
=
\frac{1}{2\pi i}
\int_{\Gamma}
dz\frac{e^{zt}}
{z+\frac{C}{N} \sum_{k=1}^N\frac{1}{\omega_k}\frac{1}{i\Delta_{\omega_k}+z}}
\label{F_{}}
\ee
where $\Gamma$ is any contour parallel to the imaginary axis and to
the right of all the poles of the integrand.  

The poles of the integrand are equal to the eigenvalues of
\be
\left(
\begin{array}{ccccc}
0 &
-\sqrt{\frac{C}{\omega_1 N}} 
&
\dots
&

&
-\sqrt{\frac{C}{\omega_N N}} 
\\
\sqrt{\frac{C}{\omega_1 N}} 
&i\Delta_{\omega_1}&0&\dots&0\\
\vdots& & & & \\
\sqrt{\frac{C}{\omega_N N}} 
&0& \dots &  & 
i\Delta_{\omega_N}
\end{array}
\right)
\ee
and, hence, are purely imaginary.

The probabilities we are interested in are 
\be
p(t)=\Bigg|\sum_{N=1}^\infty\sum_{\omega_1\dots\omega_N}
|O^{(N)}(\omega_1\dots\omega_N)|^2
F(t)_{\omega_1\dots\omega_N}\Bigg|^2\label{N-p(t)}
\ee
for (\ref{sp2}), and 
\be
p(t)'=\sum_{N=1}^\infty\sum_{\omega_1\dots\omega_N}
|O^{(N)}(\omega_1\dots\omega_N)|^2
|F(t)_{\omega_1\dots\omega_N}|^2\label{N-p(t)'}
\ee
for (\ref{sp3}). Of some interest is the fact that a ``monochromatic vacuum" 
with only one frequency, say $\omega$, implies 
\be
p(t)=\Big|\sum_{N=1}^\infty
|O^{(N)}(\omega\dots\omega)|^2
F(t)_{\omega\dots\omega}\Big|^2=
|F(t)_{\omega}|^2=p(t)'
\ee
where $|F(t)_{\omega}|^2$ is the solution for $N=1$. An analogous
property will hold for non-CCR quantized field in cavity quantum
electrodynamics: Having only one frequency we shall always end up
with the standard Rabi oscillations independently of the number of
oscillators used to model the field. 

To have a better insight into the meaning of the two probabilities it
is useful to discuss in more detail the two limiting cases: $N=1$ and
$N\to \infty$.  

\section{$N=1$ and Rabi oscillations}

For $N=1$ one finds the standard solutions
\be
F(t)_\omega
&=&
\frac{1}{2}
\Bigg(1-
\frac{\Delta_\omega}{\sqrt{\Delta_\omega^2+\frac{\hbar\alpha^2}{2\omega
m}}}\Bigg)e^{-\frac{i}{2}\big(\Delta_\omega+\sqrt{\Delta_\omega^2
+\frac{\hbar\alpha^2}
{2\omega m}}\big)t}\nonumber\\
&\pp =&+
\frac{1}{2}
\Bigg(1+
\frac{\Delta_\omega}{\sqrt{\Delta_\omega^2+\frac{\hbar\alpha^2}{2\omega
m}}}\Bigg)e^{-\frac{i}{2}
\big(\Delta_\omega-\sqrt{\Delta_\omega^2+\frac{\hbar\alpha^2}
{2\omega m}}\big)t}
\label{F N1}\\
|F(t)_\omega|^2
&=&
1
-
\frac{\hbar\alpha^2}
{\hbar\alpha^2+2\omega m\Delta_\omega^2}
\sin^2 \sqrt{\frac{\hbar\alpha^2}{2\omega m}
+
\Delta_\omega^2}\frac{t}{2}\label{|F| N1}
\ee
If vacuum is ``flat", i.e the probabilities are 
$|O(\omega)|^2=\rm const$ for $\omega<\omega_{\rm max}$ then
\be
\lim_{\omega_{\rm max}\to\infty}p(t)=
\lim_{\omega_{\rm max}\to\infty}p(t)'=1
\ee
One can understand this result as follows: For a flat vacuum the
emission is dominated by processes far from resonance. However, we
shall see in the next section that a correct physical interpretation
may require the use of a renormalized coupling constant, and the
above interpretation may be premature.  

Another interesting case is when one of the frequencies $\omega$
equals $\omega_0$ (exact resonance) while the remaining ones are very
far from resonance. Then 
\be
F_{\omega}(t)
&\approx&
\left\{
\begin{array}{lll}
\cos \sqrt{\frac{\hbar}{2\omega_0 m}}\frac{\alpha t}{2}, & 
{\rm for} & \omega=\omega_0\\
1 & {\rm for} & \omega\neq \omega_0
\end{array}
\right.
\ee
Physically this form is much more realistic than the ``monochromatic
vacuum" discussed at the end of the previous section. Then 
\be
p(t)'
&=&
1-|O_{\omega_0}|^2 \sin^2 \sqrt{\frac{\hbar}{2\omega_0
m}}\frac{\alpha t}{2}, \\ 
p(t)
&=&
\Big|1-|O_{\omega_0}|^2 \Big(1-\cos \sqrt{\frac{\hbar}{2\omega_0
m}}\frac{\alpha t}{2}\Big)\Big|^2 
\ee
Both expressions coincide if $|O_{\omega_0}|^2=1$. The analysis given
in the next sections will show that it is $p(t)$ and not $p(t)'$ that
agrees with experiment if $N$ is very large. We shall return to this
question.  

\section{Perturbative expansion for a large number of oscillators}

At this point we can estimate to what extent the noncanonical
formalism agrees with the usual one. Let us take the truncation
$p_{n_{\rm max}}$ of (\ref{N-p(t)}) at the perturbative order $n_{\rm
max}<\infty$.  
Assume there are exactly $N$ oscillators, $N>n_{\rm max}$, and the
vacuum is of the product form i.e.  
\be
O^{(N)}(\omega_1\dots\omega_N)=O_{\omega_1}\dots O_{\omega_N}
\ee
with $\sum_\omega |O_\omega|^2=1$. 

Using (\ref{f...f}) we obtain
\be
p_{n_{\rm max}}(t)
&=&
\Bigg|
\sum_{\omega_1\dots\omega_N}
|O_{\omega_1}|^2\dots |O_{\omega_N}|^2
\sum_{n=0}^{n_{\rm max}}(-1)^n
\frac{\alpha^{2n}\hbar^n}{(8m)^n}
\nonumber\\
&\pp =&\times
\int_0^{t}dt_1
\dots
\int_0^{t_{2n-1}}dt_{2n}
\frac{1}{N^n}
\sum_{l_1\dots l_n=1}^N\frac{e^{-i\Delta_{\omega_{l_1}}(t_1-t_2)}}
{\omega_{l_1}}\dots
\frac{e^{-i\Delta_{\omega_{l_n}}(t_{2n-1}-t_{2n})}}
{\omega_{l_n}}\Bigg|^2.\label{N>n}
\ee
The sum under the multiple integral in (\ref{N>n}) contains $N^n$
elements which are in a one-to-one relation with points of an
$n$-dimensional cube whose edges have length $N$ and which is embeded
in an $n$-dimensional cubic lattice. Let us denote by $N_1$ the
number of points in this cube whose indices are all different, by
$N_2$ the number of points which have exactly two identical indices,
and so on. Using this notation we can write 
\be
{}&{}&
p_{n_{\rm max}}(t)
=
\nonumber\\
&{}&
\Bigg|
\sum_{n=0}^{n_{\rm max}}
\Big(\frac{-\alpha^{2}\hbar}{8m}\Big)^n
\int_0^{t}dt_1
\dots
\int_0^{t_{2n-1}}dt_{2n}
\Bigg(\frac{N_1}{N^n}
\sum_{\omega_1\dots\omega_n}
|O_{\omega_1}|^2\frac{e^{i\Delta_{\omega_{1}}(t_2-t_1)}}
{\omega_{1}}
\dots
|O_{\omega_n}|^2\frac{e^{i\Delta_{\omega_{n}}(t_{2n}-t_{2n-1})}}
{\omega_{n}}\nonumber\\
&\pp =&\pp{====================}
+\dots +
\frac{N_n}{N^n}
\sum_{\omega_1}
|O_{\omega_1}|^2
\frac{e^{-i\Delta_{\omega_{1}}(t_1-t_2+\dots+t_{2n-1}-t_{2n})}}
{\omega_{1}^n}
\Bigg)\Bigg|^2.\nonumber
\ee
It is clear that the last term involves $N_n=N$ and the coefficient
$N_n/N^n=1/N^{n-1}\to 0$ with $N\to \infty$ (and $n>1$). A geometric
argument shows that the same holds for any $k>1$, i.e.  
$N_k/N^n\to 0$, while $N_1/N^n\to 1$. Indeed, the limit $N\to \infty$
(meaning that one increases the length of the cube's edge while
keeping the distance between the lattice points constant) is
geometrically equivalent to the limit where one keeps the length of
the edge fixed (say 1) and increases the number of lattice points in
$[0,1]^n$. Then $\lim_{N\to\infty}N_k/N^n$, $k>1$, is the probability
of finding the point $(x_1,\dots,x_n)\in [0,1]^n$ whose $k$
coordinates are equal. Sets of such points have $n$-dimensional
measure zero, as sets of geometric dimension at most $n-1$ (think of
probability of hitting a diagonal in a square).  

It follows that for a sufficiently large $N$ one can keep only the
first term i.e.  
\be
{}&{}&
p_{n_{\rm max}}(t)
\approx\nonumber\\
&{}&
\Bigg|
\sum_{n=0}^{n_{\rm max}}(-1)^n
\frac{\alpha^{2n}\hbar^n}{(8m)^n}
\int_0^{t}dt_1
\dots
\int_0^{t_{2n-1}}dt_{2n}
\sum_{\omega_1\dots\omega_n}
|O_{\omega_1}|^2\frac{e^{i\Delta_{\omega_{1}}(t_2-t_1)}}
{\omega_{1}}
\dots
|O_{\omega_n}|^2\frac{e^{i\Delta_{\omega_{n}}(t_{2n}-t_{2n-1})}}
{\omega_{n}}
\Bigg|^2\nonumber
\ee
In the Appendix we show that the right-hand-side of this expression
coincides with the perturbative expansion of the survival amplitude
in the standard canonical theory, truncated at $n_{\rm max}$, whose
Hamiltonian is 
\be
H_{\rm reg}&=&
\hbar\omega_0 R_3 +\sum_\omega\hbar\omega a{_\omega}^{\dag}a{_\omega}
+i\frac{\hbar\alpha}{2} 
\sum_\omega |O_\omega |
\sqrt{\frac{\hbar}{2\omega m}}\Big(
R_+a{_\omega}
-
R_-a{_\omega}^{\dag}\Big)
\ee
and $a{_\omega}$, $a{_\omega}^{\dag}$ satisfy the CCR algebra with
unique vacuum. Square summability of $O_\omega$ implies $|O_\omega
|\to 0$ with $\omega\to\infty$. It is clear that $H_{\rm reg}$ is the
standard RWA Hamiltonian with cut-off functions $|O_\omega |$. It is
important that the non-CCR formalism introduces the cut-off
automatically.  A difference with respect to the usual ad hoc
regularizations is that the cut-off functions are here summable to
unity and, hence, cannot equal 1. Assume
that $|O_\omega |=A=\rm const$ until the cut-off region and then
decay to 0 in order 
to guarantee $\sum_\omega |O_\omega |^2=1$. The regularized
formulas of the canonical theory agree with the non-canonical ones if
one redefines the coupling constant by $\alpha_{\rm exp}=A\alpha$.
The parameter $\alpha$ plays therfore a role of a bare coupling
constant \cite{I}.

\section{Nonperturbative amplitude for a large number of oscillators}

The parameters $C$, $N$ and $\omega_1,\dots\omega_N$ in (\ref{F_{}})
are fixed, integration is 
over any contour localized to the right of all the poles, and the poles are 
imaginary. It follows that 
the contour can be shifted sufficiently far to the right so that the
inequality 
\be
\Big|\frac{C}{N} 
\sum_{k=1}^N\frac{1}{\omega_k}\frac{1}{z}\frac{1}{i\Delta_{\omega_k}+z}\Big|<1
\ee
is satisfied and
\be
\frac{1}{1+\frac{C}{N} 
\sum_{k=1}^N\frac{1}{\omega_k}\frac{1}{z}\frac{1}{i\Delta_{\omega_k}+z}}
&=&
\sum_{n=0}^\infty \Big(
-\frac{C}{N} 
\sum_{k=1}^N\frac{1}{\omega_k}\frac{1}{z}\frac{1}{i\Delta_{\omega_k}+z}\Big)^n
\ee
The amplitude of interest can be thus written as 
\be
F(t)_{\omega_1\dots\omega_N}
&=&
\frac{1}{2\pi i}
\sum_{n=0}^\infty (
-C)^n
\int_\Gamma
dz\frac{e^{zt}}{z^{n+1}}
\frac{1}{N^n}
\Big(\sum_{k=1}^N\frac{1}{\omega_k}\frac{1}{i\Delta_{\omega_k}+z}\Big)^n
\ee
Repeating the argument from the previous section one can show that
for a sufficiently large $N$ 
\be
\sum_{\omega_1\dots\omega_N}
|O_{\omega_1}|^2\dots|O_{\omega_N}|^2
F(t)_{\omega_1\dots\omega_N}
\approx
\frac{1}{2\pi i}
\int_{\Gamma}
dz\frac{e^{zt}}
{z+C\sum_\omega |O_\omega|^2\frac{1}{\omega}\frac{1}{i\Delta_{\omega}+z}}
\ee
For a large $N$ one expects therefore the same expressions as in the
canonical case (cf. the Appendix), but with the coupling constant
appropriately regularized by the ``vacuum form-factors" $|O_\omega|$.

\section{Vacuum energy and the cosmological constant problem}

In standard canonical quantum field theories the vacuum contribution
to energy density is \cite{Milloni} 
\be
\rho_{\rm old}
&=&
\frac{E}{V}
=\frac{1}{V}\sum_k\frac{\hbar\omega_k}{2}
\approx
\frac{\hbar}{4\pi^2c^3}\int_0^{\omega_{\rm max}}\omega^3d\omega
=
\frac{\hbar}{16\pi^2c^3}\omega_{\rm max}^4
\ee
where $V$ is the volume and $\omega_{\rm max}$ a cut-off frequency. 
The choice of a concrete value of $\omega_{\rm max}$ is a question of taste. 
One of the typical candidates for $\omega_{\rm max}$ is the frequency
associated with the Planck scale. The problem is that the resulting
estimate on the value of cosmological constant is some $10^{120}$
times too big when compared with
experiments\cite{Weinberg,Witen,Rugh}.  

The energy of vacuum in our model of a bosonic quantum field is 
\be
\langle \uu O|\uu H|\uu O\rangle
&=&
\frac{\hbar}{2}\langle \uu O|\uu \Omega|\uu O\rangle
\ee
The resulting vacuum energy density is 
\be
\rho_{\rm new}
&=&
\frac{\langle N\rangle}{V}\sum_k\frac{\hbar\omega_k}{2}|O_\omega|^2
\approx
\langle N\rangle
\frac{\hbar}{4\pi^2c^3}\int_0^\infty\omega^3|O(\omega)|^2d\omega
\ee
where $\langle N\rangle$ is the average number of oscillators and we
have approximated the discrete probabilities $|O_\omega|^2$ 
by an appropriate probability density 
$|O(\omega)|^2$ normalized by 
\be
\sum_k|O_{\omega_k}|^2
=
\frac{V}{2\pi^2c^3}\int_0^\infty|O(\omega)|^2\omega^2d\omega
=1
\ee
Assuming for simplicity that $|O(\omega)|^2$ is constant up to some
$\omega_{\rm max}$ and zero for $\omega>\omega_{\rm max}$ one finds 
\be
\rho_{\rm new}
\approx
\frac{3}{8}\hbar\omega_{\rm max}\frac{\langle N\rangle}{V}
\ee
Assuming further that $\omega_{\rm max}=10^a$s$^{-1}$ and comparing
$\rho_{\rm new}$ with the experimental value $\sim 10^{-47}$GeV$^4$
one gets $\langle N\rangle/V\sim 10^{18-a}$cm$^{-3}$, which is the
average number of oscillators per cubic centimeter in the universe in
our toy model of quantum field theory. To compare with atomic data
one should not have the cut-off at wavelengths shorter than the Bohr
radius, corresponding to $\omega_{\rm max}\sim 10^{18}$s$^{-1}$,
which yields roughly one oscillator per cm$^3$, or even less. Such a
result seems to contradict the idea that the number $N$ of
oscillators interecting with atomic electrons is large. More reliable
estimates could be derived if one compared the cosmological constant
with predictions of non-canonically quantized fields with broken
supersymmetry.  
The analysis shows at least that the ``cosmological constant problem"
has to be formulated in different terms if non-canonical description
of quantum fields is employed.

\section{Fock bundle --- the non-canonical space of states}

The canonical theory has served as a kind of a reference frame for
our non-canonical calculations. An important conclusion is that the
non-canonical survival probabilities which, for large $N$, coincide
with predictions of the canonical theory have to be computed as if
the set of vacuum states was one dimensional: Survival
probabilities, denoted $p(t)$, are averages of the projector $|\uu
O\rangle\langle \uu O|$, in exact analogy to the standard CCR formalism. 

However, the set of vacuum states is infinite dimensional: There
are infinitely many different states annihilated by $\uu a_\Omega$. 
Survival probabilities computed as averages of the projector
projecting on the entire subspace of vacuum states, denoted $p(t)'$,
typically differ from $p(t)$ and are not expected to agree with
experiments. 

The ensemble of oscillators forms an object whose properties allow us
to regard it as a bosonic (spin-0) quantum field. 
Non-CCR vacua are Bose-Einstein  condensates of the ensemble. 
States of Bose-Einstein condensates are not unique even if one deals
with bosons 
of the same type (say, sodium atoms) at zero temperature \cite{BE}. 

With each vacuum state one can associate a Fock space of all its
excitations. The Fock space is obtained in the standard way by means
of the non-CCR algebra of creation and annihilation operators. 
The set of states has here a structure of a fiber bundle. The base
space $B$ is the set of all the vacua: $|\uu O\rangle\in B$ if $\uu
a_\Omega|\uu O\rangle=0$. A fiber at $|\uu O\rangle$ is the Hilbert
space $\uu {\cal H}_{|\uu O\rangle}$ spanned by vectors of
the form $\uu a_{\omega_1}^{\dag}\dots \uu a_{\omega_n}^{\dag} |\uu
O\rangle$, for all $n=0,1,2\dots$ and all $n$-tuples
$(\omega_1,\dots,\omega_n)$. The perturbative expansion of the
vacuum-picture state vector
$e^{-i\uu H_{\rm rwa}t/\hbar}|\uu O\rangle$ consists of vectors belonging to  
$\uu {\cal H}_{|\uu O\rangle}$, and the dynamics defines a flow
inside of the fiber. Probability $p(t)$ is the survival probability
{\it in the fiber\/}. The Schr\"odinger picture may be regarded as a
description with ``moving fibers". Transition from Schr\"odinger's to vacuum
picture removes the part of dynamics in the bundle, namely the motion
of a fiber along the trajectory in $B$ caused by zero-energy part
of the Hamiltonian. The vacuum picture allows us to work only with
the part of the dynamics which is internal to a given fiber.

In a separate paper, when a relativistic non-CCR formalism will be
introduced, we shall see that the action of the Poincar\'e group on
the Fock bundle makes vacua covariant and not invariant \cite{PG}. 

\section{Discussion}

The discussion of non-canonical quantum optics presented in \cite{I}
followed the standard route outlined by Dirac \cite{Dirac}: One
starts with classical fields and replaces amplitudes by operators. In
the present paper we have reversed the logic of the construction. We
have started with a single oscillator and then showed that an
ensemble of such oscillators has properties of a non-canonical bosonic
quantum field. In order to control physics underlying the formulas we
have purposefully restricted the discussion to
nonrelativistic oscillators with mass $m$. 

In spite of this restriction the formalism we arrive at has
properties strikingly similar to those of radiation fields. 
The main similarities are found in perturbative formulas describing
interactions of such fields with two-level systems. They include Rabi
oscillations and spontaneous emission. 
Having fixed the order of perturbation theory we can always find a
finite $N$ which gives predictions equal to the canonical ones
within a given precision. To put it differently, for sufficiently
large but finite $N$ one expects differences with respect to the
canonical theory to be seen in long-time tails.

The main consequence of replacing the parameter $\omega$ by an operator 
is the {\it automatic\/} disappearance of both ultraviolet and vacuum
infinities. It is 
therefore not excluded that the missing element of contemporary
quantum field theories is that they are not quantized {\it enough\/} and are
conceptually rooted in the 1925 matrix version \cite{1925} of the
old-fashioned quantum theory.  

The comparison of perturbative calculations in canonical and
non-canonical theories suggests that the structure of the space of
states appropriate for quantum field theory is not that of the Fock
one with unique vacuum, but rather of a vector bundle with all the
possible vacua in the role of a base space with Fock-type fibers. 

The disappearance of the zero-energy infinity sheds new light on the
cosmological constant problem. A natural guess is that an extension
of non-canonical methods to superfields with broken supersymmetry is
necessary in order to make cosmological predictions more realistic.
Another element which needs consideration is the question of locality
of non-CCR-quantized fields. It is quite evident that such fields
cannot be local if $|O_\omega|\neq \rm const$. 

These are natural directions for further investigations.

\acknowledgments
The work of MC was supported by the Alexander von Humboldt Foundation and
NATO. MC is indebted to prof. H.-D. Doebner and Jan Naudts for
hospitality in Clausthal and Antwerp. Many interesting
discussions with H.-D. Doebner, Wolfgang L\"ucke, J. Naudts, and
Maciek Kuna are gratfuly acknowledged.

\section{Appendix: Comparison with the canonical case}

We start with the interaction-picture Hamiltonian 
\be
H_{\rm reg}(t)&=&
i\frac{\hbar\alpha}{2} 
\sum_\omega |O_\omega |
\sqrt{\frac{\hbar}{2\omega m}}\Big(
R_+a{_\omega}e^{-i\Delta_\omega t}
-
R_-a{_\omega}^{\dag}e^{i\Delta_\omega t}
\Big)
\ee
and the initial state $|O_{\rm ccr}\rangle=|0_{\rm ccr}\rangle|+\rangle$. 
Creation and annihilation operators satisfy the CCR algebra 
$[a_\omega,a_{\omega'}^{\dag}]=\delta_{\omega\omega'}$. 
Since $R_+=|+\rangle\langle -|$, $R_-=|-\rangle\langle +|$, one has 
$R_\pm^2=0$, $\langle +|R_-=0$, $R_+|+\rangle=0$,
$R_+R_-=|+\rangle\langle +|=(R_+R_-)^n$, $(R_+R_-)^nR_+=R_+$, 
$\langle +|(R_+R_-)^nR_\pm|+\rangle=0$. 

The perturbative expansion of the survival amplitude (up to an
overall phase factor) is 
\be
F(t)&=&\langle O_{\rm ccr}|
\sum_{n=0}^\infty\Big( \frac{1}{i\hbar}\Big)^n
\int_0^{t}dt_1\int_0^{t_1}dt_2
\dots
\int_0^{t_{n-1}}dt_n
H_{\rm reg}(t_1)
H_{\rm reg}(t_2)
\dots 
H_{\rm reg}(t_n)
|O_{\rm ccr}\rangle\\
{}&=&
\sum_{k=0}^\infty
(-1)^k\frac{\alpha^{2k}\hbar^k}{8^km^k}
\int_0^{t}dt_1\int_0^{t_1}dt_2
\dots
\int_0^{t_{2k-1}}dt_{2k}
\sum_{\omega_1\dots\omega_{2k}}
|O_{\omega_1}|\dots |O_{\omega_{2k}}|
\sqrt{\frac{1}{\omega_1}}\dots \sqrt{\frac{1}{\omega_{2k}}}
\nonumber\\
&\pp =&\times
\langle 0_{\rm ccr}|
a_{\omega_1}e^{-i\Delta_{\omega_1}t_1}
a_{\omega_2}^{\dag}e^{i\Delta_{\omega_2}t_2}
\dots 
a_{\omega_{2k-1}}e^{-i\Delta_{\omega_{2k-1}}t_{2k-1}}
a_{\omega_{2k}}^{\dag}e^{i\Delta_{\omega_{2k}}t_{2k}}
|0_{\rm ccr}\rangle\nonumber\\
{}&=&
\sum_{n=0}^\infty
(-1)^n\frac{\alpha^{2n}\hbar^n}{(8m)^n}
\int_0^{t}dt_1\int_0^{t_1}dt_2
\dots
\int_0^{t_{2n-1}}dt_{2n}
f_O(t_1-t_2)
\dots 
f_O(t_{2n-1}-t_{2n})\label{A1}\\
&=&1-\frac{\alpha^{2}\hbar}{8m}
\int_0^{t}dt_1\int_0^{t_1}dt_2
f_O(t_1-t_2)F(t_2)\label{A2}
\ee
where 
\be
f_O(\tau)=\sum_\omega|O_\omega|^2\frac{e^{-i\Delta_\omega\tau}}{\omega}
\label{fOtau}. 
\ee
The survival amplitude is 
\be
F(t)
&=&
\frac{1}{2\pi i}
\int_\Gamma
dz\frac{e^{zt}}{z+C 
\sum_{\omega}|O_\omega|^2\frac{1}{\omega_k}\frac{1}{i\Delta_{\omega_k}+z}}
\label{A3}
\ee
Although there are evident similarities between the latter formula
and (\ref{F_{}}), some differences are also very interesting. Of
particular importance is the fact that in (\ref{F_{}}) the
frequencies are fixed and the sum is over the number of oscillators.
The canonical case (\ref{A3}) involves the sum over all frequencies.
Similar differences are between (\ref{fOtau}) and (\ref{ftau}).  


\begin{references}
\bibitem{footnote}Let us try to imagine such an oscillator and what
it may mean that 
$\omega$ is a quantum number. Let us first recall that for a simple
(classical) pendulum in the linear approximation $\omega=\sqrt{g/l}$
where $g$ is the gravitational acceleration and $l$ the length of the
pendulum. A simplest ``simple quantum pendulum" may be imagined as a
two-atom molecule with one of the atoms trapped in some ``ceiling"
and the other ``hanging down". This is an idealization, of course, but
not worse than other idealizations one often makes in quantum
mechanics. The 
role of the length $l$ is played by the distance between centers of
mass of the two atoms. Since the two atoms are described by
configuration-space wavepackets, they are in superpositions of
different center-of-mass localizations. Accordingly, the pendulum is
in a superposition of different lengths and, hence, also of
$\omega$s. Frequency operator becomes 
$\Omega=\sqrt{g/|r_1-r_2|}$ where $r_k$ are
center-of-mass operators of the atoms. The limiting values of
spectrum of $\Omega$ are $\omega=0$ (if the length is infinite) and
$\omega=\infty$ (if the length is zero). For a molecule both
conditions can be assumed to occur with zero probability. It follows
that the wave function should vanish at $\omega=0$ and
$\omega=\infty$. There exists an analogue of these properties for
Fourier transforms of massless fields \cite{infrared}. 

One may criticize the example since such an $\Omega$ does not
commute with canonical center-of-mass momenta of the atoms. This is
true indeed but what it shows is that superpositions of $\omega$s
are quite natural if quantum oscillators are considered.
\bibitem{infrared}Ultraviolet infinity is automatically regularized in this
formalism by square-integrability of vacuum states since probability
of high four-momenta tends to zero. The same property makes vacuum
fluctuations finite. It is known that momentum-space wave functions
of massless fields are differentiable functions vanishing at the
origin $p=0$ of the light cone 
\cite{Woodhouse} a property which is related to the structure of
unitary representations of the Poincar\'e group. One expects that
the latter feature will regularize the infrared infinities. The
problem requires relativistic considerations and is beyond
the scope of the present paper.
\bibitem{hat 1}To avoid proliferation of different symbols we denote
the two (in general different) identity operators in $\hat
1\otimes\hat 1$ by the same symbol --- this will not lead to
ambiguities.
\bibitem{BE}Bose-Einstein  condensates of sodium atoms created at two
different laboratories are different since their position-space wave
functions are localized at different places. Excitations (phonons) in such 
condensates can be regarded as elements of different Fock spaces.
Each state space is a fiber in a Fock bundle whose base space is the set of
all the possible vacua.
\end{references}
\end{document}